*agronomy* 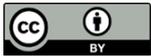

*Article*

# Bibliometric and social network analysis on the use of satellite imagery in agriculture: an entropy-based approach

Riccardo Dainelli [1,*] and Fabio Saracco [2]

1   National Research Council - Institute of BioEconomy (CNR-IBE), Via Caproni 8, 50145 Firenze, Italy
    riccardo.dainelli@ibe.cnr.it
2   "Enrico Fermi" Research Center (CREF), Via Panisperna 89A, 00184 Rome, Italy; fabio.saracco@cref.it
*   Correspondence: riccardo.dainelli@ibe.cnr.it

**Abstract:** Satellite imagery is gaining popularity as a valuable tool to lower the impact on natural resources and increase profits for farmers. The purpose of this study is twofold: to mine the scientific literature for revealing the structure of this research domain and to investigate to what extent scientific results are able to reach a wider public. To fulfil these, respectively, a Web of Science and a Twitter dataset were retrieved and analysed. Regarding academic literature, different performances of the various countries were observed: the USA and China resulted as the leading actors, both in terms of published papers and employed researchers. Among the categorised keywords, "resolution", "Landsat", "yield", "wheat" and "multispectral" are the most used. Then, analysing the semantic network of the words used in the various abstracts, the different facets of the research in satellite remote sensing were detected. It emerged the importance of retrieving meteorological parameters through remote sensing and the broad use of vegetation indexes. As emerging topics, classification tasks for land use assessment and crop recognition stand out, together with the use of hyperspectral sensors. Regarding the interaction of academia with the public, the analysis showed that it is practically absent on Twitter: most of the activity therein is due to private companies advertising their business. Therefore, there is still a communication gap between academia and actors from other societal sectors.

## 1. Introduction

*1.1 A help from above: the use of satellite imagery in agriculture*

In sustainable agriculture, the necessity of having lower exploitation of agronomic inputs and natural resources must be reconciled with high production standards, increasing profits for farmers [1] and providing social benefits [2]. In this context, satellite platforms play an important role in two main application branches: agricultural machinery geolocation and navigation and field remote sensing (RS). In the first case, satellites dedicated to the Global Navigation Satellite System (GNSS) permits, among applications, in-field georeferenced observations, Variable Rate Technology (VRT) treatments and the automation of tractor-driving systems [3,4]. Satellite remote sensing currently includes numerous constellations capable of providing a wide range of imagery products which Precision Agriculture (PA) makes use of in many different applications [5].

The launch of the NASA Landsat-1 satellite in 1972 represents the first step in the evolution of satellite remote sensing for applications in agriculture [6]. The year 2008, on the other hand, marks the beginning of the era of microsatellites and satellite constellations: from here on, it was possible to obtain usable and accurate imagery with a high acquisition frequency and covering large areas [7]. Finally, nowadays, as well as national space agencies, private companies are launching their agriculture-focused satellites [8].

Satellites provide information by measuring the electromagnetic radiation reflected or emitted from the earth's surface being equipped with a wide range of sensors that allow spatial and multitemporal monitoring of soil and crop features throughout all the phenological stages. Such sensors can be active or passive [5]. Active sensors, radar and lidar, use an artificial source of energy to illuminate objects on the ground and detect their reflected response or backscattering. Passive sensors instead detect the amount of radiation reflected or emitted by bodies in specific intervals of the spectrum e.g. the visible region (VIS, 380-750 nm), the near-infrared (NIR, 0.7-1.3 μm), the short infrared (SWIR, 1.3-2.5 μm), the thermal infrared (TIR, 7-20 μm) [9]. Each type of sensor has a typical application in field monitoring. Radar and lidar can be used for soil moisture estimation or the evaluation of vegetative growth. Panchromatic sensors, synthesising the red, green and blue (RGB) bands, can produce images with a high spatial resolution for drafting maps or the construction of the digital terrain model (DEM) [10]. DEMs can indeed be generated from 2D-RGB images, but this requires some specific characteristics of the imagery (e.g. sufficient overlap between two adjacent images). A typical processing approach can be performed through photogrammetric techniques, such as the so-called Structure from Motion (SfM) [11]. Multispectral sensors record several spectral bands (typically 3-12) with a high amplitude (15-90 nm), useful for the calculation of traditional vegetation indices, such as the Normalized Difference Vegetation Index (NDVI), or plant species recognition [5]. Hyperspectral sensors measure the spectral response in a very high number (hundreds) of bands, with finer amplitude (5-10 nm). Thus, they can identify absorption peaks related to certain components (e.g. photosynthetic pigments) and allow the calculation of more specific indices or the retrieval of soil organic matter. Thermal sensors estimate the temperature of vegetation and soil for evaluating soil water content, detecting crop water stress and irrigation scheduling [9].

In the last years, satellite imagery has come to the fore, not only as a research tool but also as a support to many stakeholders of the agricultural supply chain. Thus, the application of this technology for several crop practices, such as fertilisation [12], irrigation [13], harvest [14], could allow sustainable management of agricultural systems. This has been also possible thanks to recently created *ad hoc* web services. Applications range from recognition of the plot boundaries [15] to those providing interoperability between drones and satellite data [16], up to decision support systems (DSS) with customizable packages [17].

In the choice of satellite products for research or technical purposes, the spatial resolution is a compromise between imagery costs and accuracy necessary for crop monitoring [10]. Extensive crops, such as cereals, can be monitored with appreciable results using intermediate spatial resolutions (5-10 m per pixel) [12], while for crops with discontinuous coverage the presence of inter-rows involves the need for a greater resolution, as in the case of vineyards. Thus, to discriminate crops without altering the information due to the co-presence within the same pixel of the canopy and bare soil/grass, a sub-metric detail (<0.1 m per pixel) is required [18]. Where it is not possible going into detail due to sensor limitations, another way to tackle with coarse resolution and mixed pixels containing various attributes can be represented by the sub-pixel classification [19]. In the same way, timely monitoring is a key issue for agricultural activities and for this reason satellite revisit time (i.e. the frequency of platform passage over the same area) can be a discriminating factor in the choice of satellite imagery [10]. Retrieving the beginning of a phenological phase or verifying field operations require high temporal resolution which has been guaranteed in the last decade by open data (e.g. Sentinel-2 with 5 days of revisit time) [20] or commercial solutions such as WorldView or Pleiades, which provide daily images [5].

*1.2 Satellite imagery for agriculture through different information sources*

The importance of the use of satellite RS in agriculture is remarkable considering the wide range of featured applications, sensors and resolutions. Understanding the main developments in the use of satellite imagery in agriculture is crucial for synthesising the existing knowledge base, advancing the line of research and making better use of this technology. In this context, literature reviews are increasingly becoming important due to the fast-growing pace of scientific production [21]. But traditional review papers focus only on the content of publications and may suffer from the subjective biases of the researchers involved in the process [22]. By contrast, bibliometrics is applying quantitative analysis and statistics to perform a systematic, objective and reproducible analysis of the unstructured and large body of information [23]. It can be used to detect the most influential authors/journals/institutions/countries, the principal research streams, the emerging and trending topics and to show a clear overview of a research topic [24]. In doing this, bibliometrics takes advantage of scientific methods such as Social Network Analysis (SNA) [25].

Two-way communication between scientists and the public should be incorporated into effective communication and scientific outreach efforts to improve not only the reputation of particular innovations but also an appropriate development by listening to audiences and tailoring innovations to their needs [26]. A virtual forum where this type of communication can be implemented is represented by online social network platforms [27]. In this sense, it could be very interesting to take a look at how the use of satellite imagery in agriculture is presented in online social networks.

*1.3 Aim of the study*

To the best of the authors' knowledge, only a few studies addressed exclusively the use of satellite platforms and sensors for agricultural applications leveraging academic databases [28,29]. Moreover, there are still no research papers that exploited informetric analysis (scientific literature + information retrieved from online social networks) for summarising the research field under investigation. Thus, the first aim of the work is to mine the scientific literature from the first applications until today to reveal the structure of this research domain. The study is based on the following research questions (RQs):

1. what are the most influential countries and journals and the key papers?
2. What is the spectrum of the terminology related to the use of satellite imagery in agriculture and what are the relations among them?
3. What are the main research streams and emerging topics?

By answering the above questions, this paper aims at providing researchers with a comprehensive and objective framework of the topic. It also supports the scientific process by giving research directions and identifying hotspots.

In addition to the bibliometric survey, the discourse on the use of satellite imagery in agriculture is investigated within one of the most popular social network platforms, such as Twitter. The choice of Twitter is due to two main reasons: it is an easy-to-use tool to spread news and scientific results and for the completeness of the available data. Two further RQs aim to broaden the study:

4. who and for what purpose talks about satellite imagery for agriculture on Twitter?
5. What are the cutting-edge dynamics and development trends discussed?

This analysis allows examining the communication and outreach of the research topic also outside academia, providing a point of view that is more oriented to actual applications.

## 2. Materials and Methods

*2.1 WOS dataset description*

2.1.1 Scientific paper search

The authors performed a comprehensive literature search using the Web of Science (WoS - Clarivate Analytics) search engine to extract peer-reviewed studies related to the use of satellite imagery in agriculture from the first applications up to the current days. The guiding approach is similar to the PRISMA methodology [30].

The query was '(TS = ("Agric*")) AND (TS = ("Satellite"))', i.e. all papers having "satellite" and all words beginning with "agric-" in the abstract. The research words have been deliberately chosen not to be too specific for encompassing all the works related to the topic without risking excluding any item. Then, a filtering step was conducted by exploiting the exclusion criteria directly available in the WoS search engine, that is document type, language and publication year. Original articles, review articles, book chapters, editorial material and data papers published in the English language were selected. Regarding the timespan, the investigation encompassed the entire body of literature from 1977 to mid-2022 (30 June). The search query generated 8526 records with publication data containing information on the "Author, Title, Source", "Abstract, Keyword, Addresses" and "Cited References and Use" categories, resulting in 72 fields. Records not relevant to the research goal (e.g. satellite RNA in plant pathology), were excluded in a data cleaning step (more details in the following sections).

2.1.2 Data cleaning

As with any dataset, WoS extraction is affected by some noise: for instance, some records do not display the abstract or others do not display the publication year. Moreover, it seldom happens that some papers that are not related to the research field under investigation are included: for instance, the terms "satellite" and "agriculture" are used in the context of archaeology (for *satellite* settlements and their *agricultural* activities) or in urban studies (for urban pattern management).

To get rid of those records that may put at risk the coherence of the dataset, the following procedure was implemented. Each record in the dataset is provided with at least a single *Research Area* (RA), i.e. the field of the research. A manual selection of the most relevant for the research was performed choosing *remote sensing, agriculture, imaging science & photographic technology, science & technology -other topics, engineering, computer science, geochemistry & geophysics, plant science, instrument & instrumentation*. To be maximally conservative, all papers having all RAs in this list were selected: the resulting dataset displayed 1933 papers.

In performing the analysis, authors observed *a posteriori* that some recurrent keywords were unrelated to the topics under analysis. Even limiting the analysis to the RAs above, still some applications to cell biology and marine science were present. Therefore, authors excluded from the set of abstracts all those including the words "dna", "rna", "clone" and "virus". To limit analysis to agriculture, "ocean", a recurrent term referring to coast monitoring and "aerosol", a frequent keyword related to atmospheric science, were also excluded. The final dataset included 1830 papers. While this approach could be considered particularly strict, further analyses prove that it minimises the presence of noise.

2.1.2 Twitter dataset description

Messages published on Twitter focusing on the same topics were analysed. In particular, using the GET /2/tweets/search/all API [31] with Academic Access [32], authors accessed all messages containing the keywords used for retrieving the WoS dataset, from the very first message of Twitter in 2006 until the 1st of July 2022, mirroring the WoS query. More in detail, the query was "(agriculture OR agricultural OR agriculturalist OR agriculturalist OR agriculturalist OR agriculturalists OR agronomy OR agronomics OR agronomic OR agronomist OR agronomists) satellite".

The result was a dataset made of 57638 messages. Again, to mirror the analysis performed on the WoS dataset, all messages containing terms referring to "dna", "rna", "clone", "virus", "ocean" and "aerosol" were excluded, resulting in 55768 different tweets, posted by 42397 different accounts, as identified by their unique author id.

*2.2 Data analysis*

The analysis of the WoS dataset was divided into two main parts: the first one of descriptive statistics, which highlighted interesting trends of the production in the field of satellite imagery in agriculture and the relevance of scientific production of various countries and researchers; the second one was dedicated to the analysis of the semantic network, as obtained from the abstract of the papers. While the former is more straightforward, the main steps of the methodology of the latter will be introduced in the following sections.

Since just limited information regarding the authors of posts on Twitter is available, most of the first part of the WoS analysis was neglected on the Twitter dataset. Nevertheless, the analysis of semantic networks was performed on Twitter's new content messages, i.e. those that are not retweets. It is worth mentioning that this approach is similar, but different, to others in the literature. For instance, in [33] the semantic network was inferred from tweets' hashtags only.

The procedure used to get the semantic networks is pictorially represented in Figure 1. While Figure 1 shows the procedure for WoS, the one used for Twitter is perfectly analogous, just replacing WoS ids with Tweet ids and the abstracts with tweets' texts.

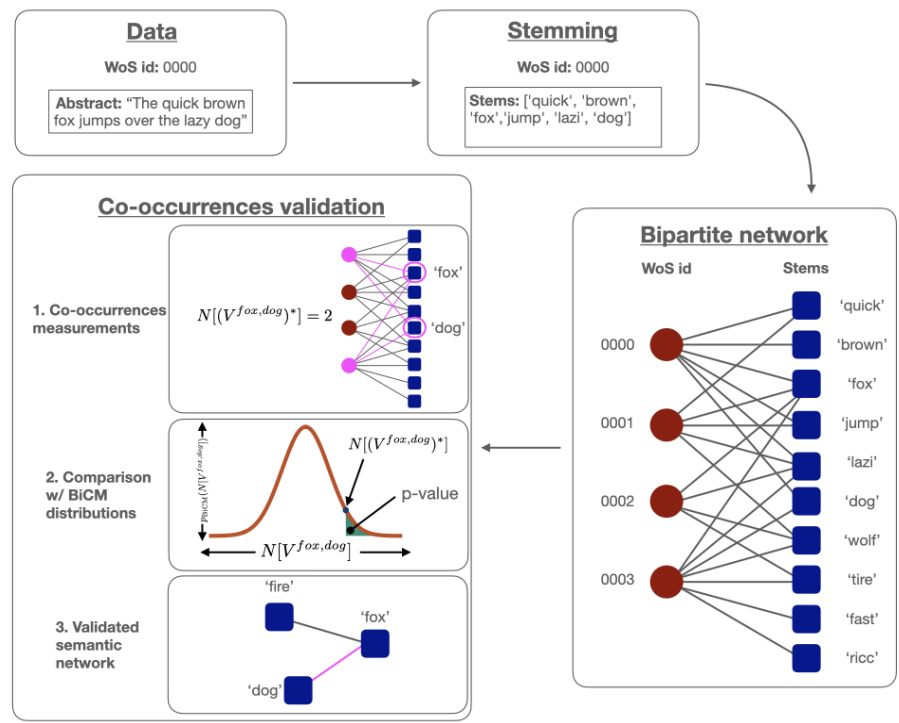

**Figure 1**. Pictorial representation of the procedure for inferring a semantic network from WoS data.

First, the text of each abstract is considered and divided into stems, using `nltk` Python module [27]. Then a bipartite network of abstracts, as identified by WoS ids, and stems, is built. The information contained in the bipartite network is subsequently projected into the layer of stems and finally validated: the projection process is obtained by comparing the value measured in the real networks with the distributions expected by a maximum-entropy benchmark. In the following, each step of this procedure is going to be described in detail.

2.2.1 Stemming words from the abstracts and tweets

All the various texts - may they be scientific paper abstracts or Twitter posts - were automatically analysed using the Natural Language Processing (NLP) tools provided by the `nltk` Python module [34]. At first, all single words were detected by removing punctuations, numbers (still considering alpha-numeric entries) and stop words. The (English) stop words list used is the one present in the `nltk` submodule `nltk.corpus` and includes pronouns, demonstrative adjectives, prepositions, auxiliary verbs, and similar items [34].

Then, the Snowball stemmer [34] was used. A stemmer reduces a word to a "stem" by cutting prefixes and suffixes. The rationale is to reduce the different conjugations or declensions of the same word to the same stem, to correctly count the occurrence of the same concept. For instance, "*crop*", "*crops*" and "*cropping*" are all stemmed by Snowball procedure to the same stem "*crop*".

During the analysis, the stem association performed by `nltk` was inspected by manually checking the most frequent stems. In detail, the 100 most frequent terms were examined: for each term, the 5 following and 5 preceding ones in alphabetic order (i.e. the most similar ones) were checked using edit distance. In a nutshell, the edit distance counts how many characters have to be removed or substituted to pass from a word to the following one [34]. Finally, the authors manually verified if `nltk` missed a possible association between the more frequent word and the most similar ones, selecting couples of stems for which the edit distance is lower than 3 (i.e. selecting couples in which the two stems differ for a maximum of 2 characters).

The test procedure was applied to 1000 couples of stems. Less than 1% of the comparisons resulted in being not correctly recognized by `nltk`. The impact of missed association was so limited that the `nltk` procedure was considered reliable. Due to the great number of stem couples to be verified, the result from `nltk` was considered as provided by the python module. When, in the following, an incorrect association is found, it will be stressed and commented on.

Finally, each text was associated with the (filtered) stems of the words in it: in the case of WoS, each abstract was identified by the relative paper WoS id; in the case of Twitter each message was identified by its tweet id.

2.2.2 Bipartite networks, entropy-based null-models and validated semantic networks

To identify non-trivial similarities in the way concepts and ideas are used in the various abstracts, the association between research papers and stems was represented as a bipartite network [35]. In a bipartite network, nodes are divided into two sets, called *layers*, and edges can only connect nodes from different layers. Papers and stems were described as the two layers of a bipartite network: an edge connecting a stem to a paper was present if the stem was used in the paper's abstract.

The target of the analysis is to infer non-trivial similarities among stems, given their usage in the various corpora of texts. The main intuition is that two stems (and, of course, the related words) are related if they are both used in many scientific abstracts. Therefore, for every couple of stems, the number of abstracts in which they both appear was counted; in terms of the bipartite networks are called co-occurrences.

Nevertheless, a high value of the co-occurrences between two nodes on the same layer may be due only to the presence of two frequent stems within particularly verbose abstracts. Therefore, a proper benchmark for the analysis of the observed co-occurrences is needed.

In the study, an Information Theory technique to discount the information coming from the frequency of words and the number of stems in each abstract (see [36] for a review) was used.

In a nutshell, the idea is to start from a real network and define the set of all possible graphs having the same number of nodes and all possible link configurations (the ensemble). The quantity of uncertainty of the system is accounted for by the function named *Shannon* (or *Information*) *entropy* (Equation 1):

$$S = - \sum_{G \in \Gamma} P(G) \ln \ln P(G) \tag{1}$$

where *G* is a graph in the ensemble and *P(G)* is the probability of observing the graph *G*.

The main target of this construction is to provide a maximally random benchmark, but for some network properties observed in the real system. Therefore, a constrained maximisation of the Shannon entropy associated to the system can be performed, where the constraints are exactly those defining properties observed in the real network.

Remarkably, the same approach was used in [37] to derive the canonical ensembles in Statistical Physics from Information Theory. The adaptation of this framework to complex networks permitted the extension of constraints from global - as the total energy in Statistical Physics - to local ones - as the number of connections per node [38–40].

The framework described above can be applied to bipartite networks: in this case, it is possible to constrain the number of connections (i.e. the degree) for nodes in both layers. The related model is called Bipartite Configuration Model (BiCM, [41]) and it is implemented in the `NEMtropy` Python module described in [42]

Therefore, given a real network, the observed co-occurrences can be compared with their BiCM distributions. If the BiCM probability of observing the value measured in the real network is particularly low (i.e. the relative p-value is statistically significant), then the ingredients of the BiCM cannot explain the deviation from the model and the relative co-occurrences are validated [43]. The result of the validated projection procedure is a monopartite network (i.e. a network in which nodes are all of the same nature) in which an edge represents a non-trivial similarity.

The validation procedure is present in the `NEMtropy` python module mentioned above.

## 3. Results and discussion

*3.1 WoS dataset*

3.1.1 Descriptive statistics

3.1.1.1 Country performances

Analysing the number of publications per year throughout the investigated period (1977 to mid-2022) (Figure 2), it becomes immediately evident how the number of publications has experienced rapid growth in the last years. In fact, from 2017 the number of publications breaks through the threshold of 100; the trend is further accentuated in 2022 since there are already 130 research works by mid-year.

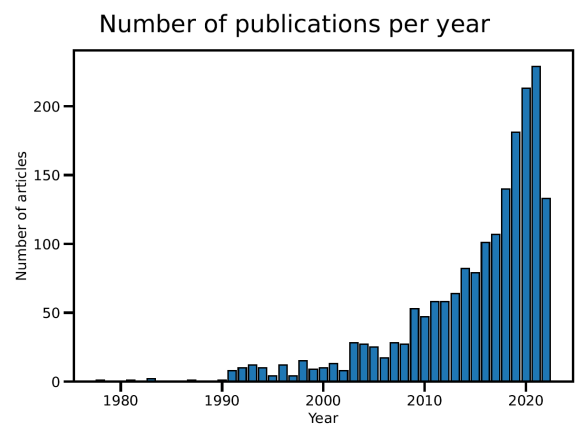

**Figure 2.** Number of publications per year (1977 - mid2022).

The most active country in the whole period encompassed by this study is the USA with over 400 publications (Figure A1). Quite far apart from the USA, the People's Republic of China (PRC) and India are positioned in the second and third place of the ranking, respectively with more than 200 and 150 research works. In a more detached group, then there are some European countries (Germany, Italy and Spain with <150 publications). Looking at the data per decade (Figure 3), the USA remained the leading country for three decades (1991-2000, 2001-2010, 2011-2020) and then was finally overtaken by PRC in 2021-2022. India ranked second in 1991-2010, then fell fourth (2011-2020) and nowadays climbs back to third place. Italy has risen from seventh to the fourth position over time while England gradually dropped out of the top 20. Other countries such as Germany and France steadily occupy ranks in the top 10 over the period. Data for the period 1977-1990 are not shown as there are a small number of publications.

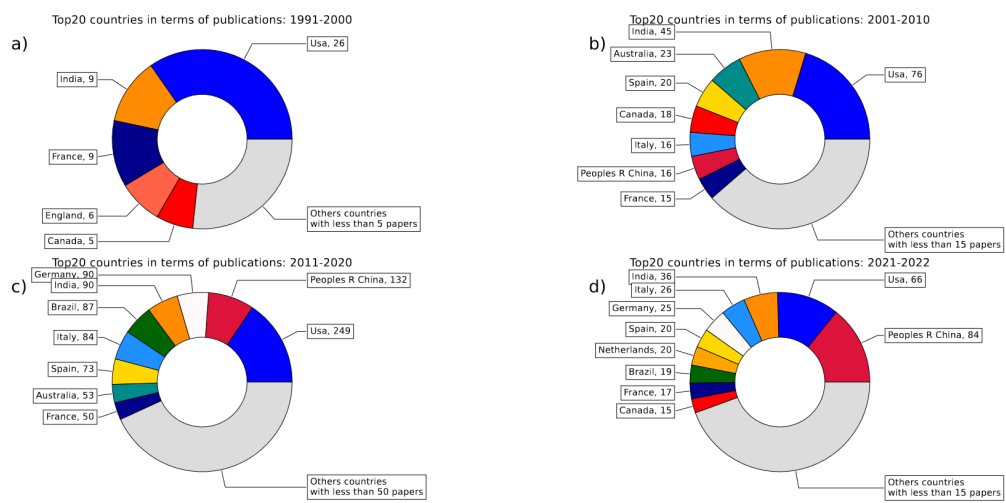

**Figure 3.** Top 20 countries in terms of publications - 1991-2000 (a), 2001-2010 (b), 2011-2020 (c), 2021-mid2022 (d).

Considering the number of researchers dealing with satellite imagery in agriculture (1977 to mid-2022) (Figure A2), the ranking mirrors that in terms of publications with the top three positions occupied by the same countries, namely the USA, PRC and India with more than 1000, 800 and 500 researchers, respectively. Brazil, Italy and Germany complete the top 6 with Italy confirming the fifth position in both rankings. Shifting from the number of researchers to the number of publications, Brazil passes from rank #4 to rank #7 while Germany from rank #6 to rank #4. This probably occurs because in Brazil more research groups are dealing with the current research topic than in Germany and/or the scientific papers are authored on average by a higher number of authors in Brazil than in Germany. Throughout the time frame of the analysis (Figure 4), the USA remains in the very first positions (#1 and #2) while a dizzying growth of PRC (out of the top 20 in 1991-2000) is evident, outperforming the USA in 2021-2022. India is also steadily in the top positions (rank #3 in 2021-2022), overcoming a slight decline in the decade 2011-2020 (rank #5). Brazil also had strong growth especially in 2011-2020 while countries like Italy, France, and Germany have always been present in the top 10. A brand-new country (Pakistan) has recently entered the top 10.

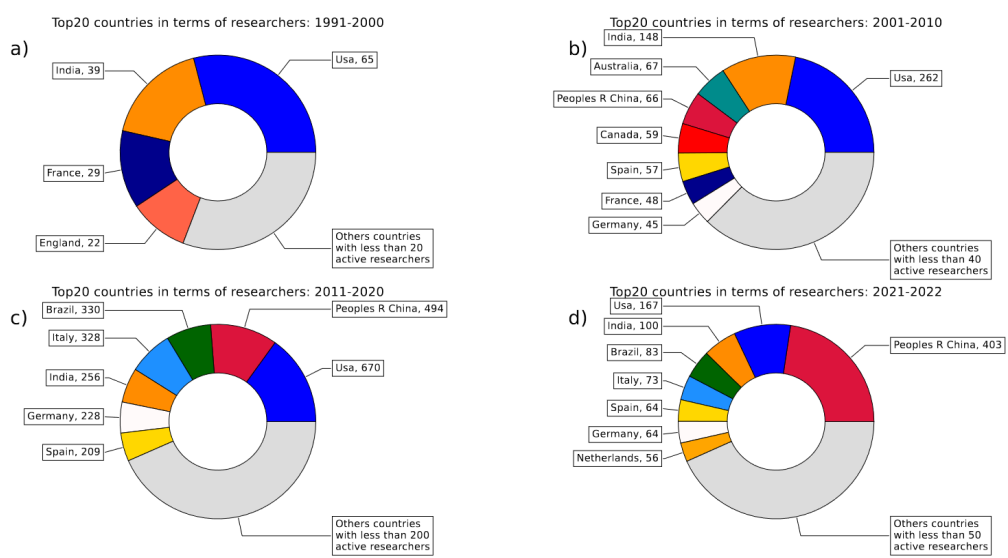

**Figure 4.** Top 20 countries in terms of researchers - 1991-2000 (a), 2001-2010 (b), 2011-2020 (c), 2021-mid2022 (d).

Nevertheless, the number of publications and researchers per country is not the end of the story. In fact, what is crucial is the impact of the research produced in each country. A proxy of the quality of research is given by the number of citations received. Such a measure is not accurate on its own, since the number of citations may be due to the institute, either the country researchers come from, or the audience (for instance, see [44]).

WoS dataset is quite limited, therefore, to have sufficient statistics, focus was placed on countries that released more than 80 papers in the entire time window (Figure 4a) and the distribution of the citations per paper was considered (Figure 4b).

In the right panel, boxplots, i.e. an intuitive way of comparing different distributions, were displayed. The edges of each box are the first (*Q1*) and the third (*Q3*) quartiles of the represented distribution, while the orange line represents the median. If the Interquartile Range is defined as *IQR = Q3 - Q1*, the whiskers are taken at *1.5 IQR* from the edges of the box. In the plot, outliers were not represented to make the plot clearer. The boxplots are ordered according to the median of each distribution.

From Figure 5, it can be observed that the impact of the research of China and India underperforms the ones of other countries, despite the strong efforts in terms of publications (see Figure 3) and researchers (see Figure 4).

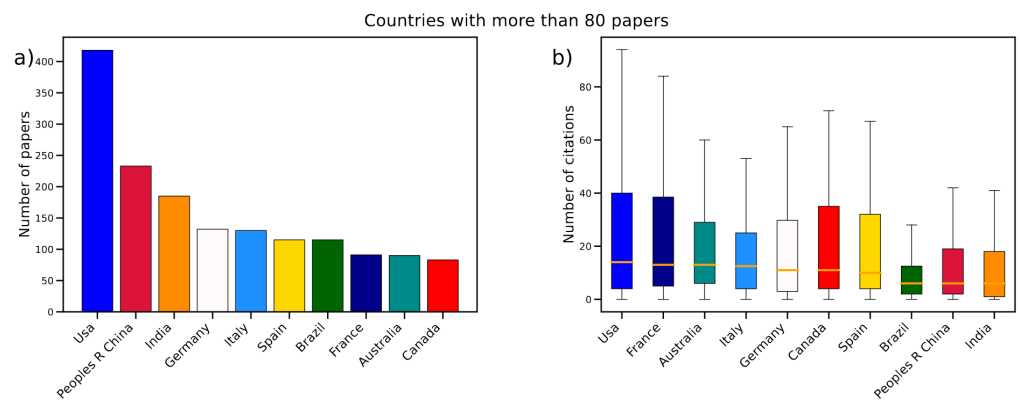

**Figure 5.** Quality of research: countries with more than 80 papers (1977 - mid2022) (a) and distribution of the citations per paper (b).

3.1.1.2 Journal performances

Among the journals that present the highest number of publications in the entire period under analysis (1977-mid2022), there are three of them that outperform all others in terms of published items (for journal abbreviation see Table S1). These are the *International Journal of Remote Sensing* (Int. J. Remote Sens. - Taylor & Francis), the *International Journal of Applied Earth Observation and Geoinformation* (Int. J. Appl. Earth Obs. Geoinf. - Elsevier) and *Computers and Electronics in Agriculture* (Comput. Electron. Agric. - Elsevier), with 307, 157 and 107 publications, respectively (Figure 6a). Taylor & Francis places only one journal in the top 10 while other publishers feature more than one as Elsevier (3, those mentioned above and *Geoderma*), MDPI (2, *Agronomy* and *Agriculture*) and Springer (2, *Precision Agriculture* and *Scientific Reports*). This picture shows that there are long-established scientific publishers but also new players who are emerging in this sector, such as MDPI and PLOS.

Focusing on the impact of each journal, the average number of citations per paper published in the journal was considered. To have sufficient statistics, only the most present journals in the WoS dataset were considered (Figure 6b). It can be immediately noted a partial reshuffling of the results reported in the top panel of Figure 6a. Indeed, the ranking of journals with the highest average number of citations is dominated by IEEE journals, that is *IEEE Geoscience and Remote Sensing Letters* and *IEEE Transactions on Geoscience and Remote Sensing* with 58.04 and 50.97 citations per paper on average, closely followed by *Agricultural Systems* (48.43 citations per paper on average). All other journals have on average a lower number of citations ranging between 38 and 5. On the contrary, few journals are positioned in the top ranking in terms of publications number but fall back many positions when analysing the average citations, i.e. *Computers and Electronics in Agriculture* and *Agronomy*. A possible explanation could lead to the specificity of the research topic covered by the journals such as agriculture (which moreover is not one of the most debated themes at the international level) and in the case of *Agronomy* due to the recent appearance of the journal on the international front of the stage (first impact factor in 2018).

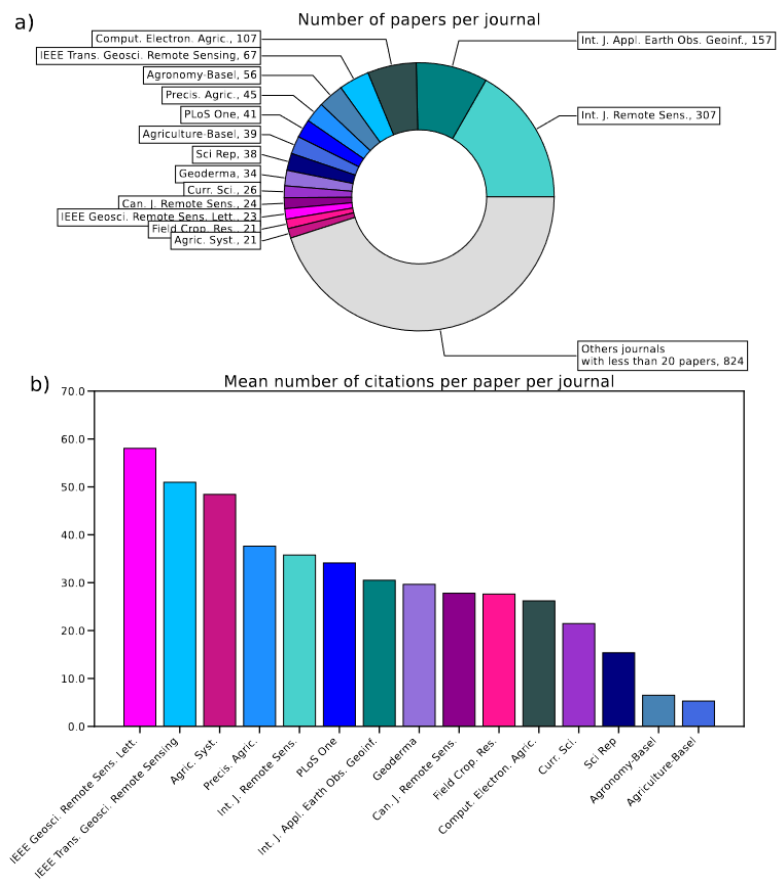

**Figure 6.** Journal performances: number of papers per journal (a) and mean number of citations per paper per journal (b).

The panels (a) and (b) of Figure 7 show, respectively, the citation's distribution within all citation indexes in WoS and the distribution of the number of papers per author. The distribution is, roughly speaking, a power law, i.e. $P(x) \sim x^{-\alpha}$. Such an observation seems to suggest that the process is a multiplicative one, i.e. if the number of publications at time *t* of a certain author is $P_t$, then its value at *t + 1* is $P_{t+1} = \varepsilon\, P_t$, where $\varepsilon$ is a certain factor [35]. In this sense, the right panel of Figure 7 seems in agreement with the dynamic suggested by Kauffman of the "adjacent possible" [45]: while the original idea derived from the context of molecular and biological evolution, it was also applied for modelling the rise of innovations (for instance, see [46]).

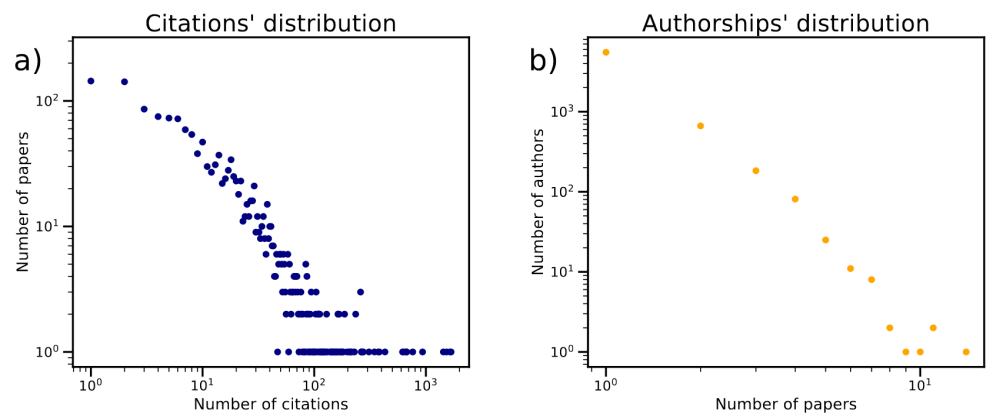

**Figure 7.** Citations' and authorships' distributions throughout the WoS dataset (1977 - mid2022).

The main idea is that innovation proceeds in single steps on a path sketched by previous contributions: in this sense, the "adjacent possible" is the set of unknown items that are at a single innovation step from the current stage of evolution. Since all new objects in the adjacent possible can be reached by recombination of existing ones, when a new item is discovered or created, the bound of the adjacent possible is updated and expanded. Modelling of such an intuition appeared in [46] and is based on power-law distributions observed in various real systems; the power-law distributions in Figure 5 support the finding therein.

The main intuition of Figure 7a is that the greatest part of the papers received a limited number of citations (left top angle), while few ones had a great number of citations (bottom right angle): analogously, in Figure 7b, a limited number of authors wrote just a paper, while a few wrote more than 10 papers on the examined subject. Finally, the power law sense is that the frequency of papers receiving *n* citations is proportional to the frequency of papers receiving $\delta_n$ via a (negative) power of $\delta$.

3.1.2 Most cited papers

Among the top 50 publications in terms of citation, 20 items strictly connected with the field of application of the study were selected. The exclusion criteria cut off studies i) on a global scale, ii) citing the search keywords < 5 times throughout the document and/or not including them in the core of the document iii) out of the aim and scope of Agronomy. The results are presented in Table 1. The most cited papers are published in the period between 1996 and 2017. If it may not be surprising that the papers published in the past decades collect many citations, for those published more recently it can be a distinctive sign of the quality of the research or at least of the novelty of the topics covered. Indeed, it is remarkable that the oldest most influential paper is relatively recent (i.e. 27 years ago). In the scientific literature, it is often stressed the importance of the first mover advantage, i.e. the fact that older papers get more citations just because they are present in the literature for a greater amount of time [47]. In this case, the situation is different: the most cited paper is only 10 years old (in a dataset 45 years long). Such an effect is probably due to the rapid evolution of the available satellite technology that may have resulted in making older papers outdated quite soon.

Among the most recent (2017), it is worth mentioning the work by Kussul et al. [48] dealing with the classification of crop type through deep learning techniques that gained 665 citations. In addition to this study, the topic of regional-level crop classification or monitoring and the land use changes towards cropland [49–52] is the backbone of the most cited paper group. The retrieving of soil properties and soil carbon seems to be another trending topic [53,54] while, despite the citations (674), the study by Scanlon et al. [55] on irrigation sustainability is the only one dealing with soil water retrieving. Regarding crop type, Ren et al. [56] estimate winter wheat yield with MODIS-NDVI data and Johnson et al. [57] exploit multispectral image to map the leaf area in a vineyard.

Contrary to what one might expect, review studies are only three among the top 20 cited publications [58–60].

**Table 1.** Top 20 papers in terms of citations. Authors, year, title and journal are also reported.

| Agricultural ranking | Authors | Year | Title | Journal | Citations |
|---|---|---|---|---|---|
| 1 | Scanlon et al. | 2012 | Groundwater depletion and sustainability of irrigation in the US High Plains and Central Valley | PNAS | 674 |
| 2 | Kussul et al. | 2017 | Deep Learning Classification of Land Cover and Crop Types Using Remote Sensing Data | IEEE Geosci. Remote Sens. Lett. | 665 |
| 3 | Morton et al. | 2006 | Cropland expansion changes deforestation dynamics in the southern Brazilian Amazon | PNAS | 618 |
| 4 | Macedo et al. | 2012 | Decoupling of deforestation and soy production in the southern Amazon during the late 2000s | PNAS | 374 |
| 5 | Gomez et al. | 2008 | Soil organic carbon prediction by hyperspectral remote sensing and field vis-NIR spectroscopy: An Australian case study | Geoderma | 346 |
| 6 | Solberg et al. | 1996 | A Markov random field model for classification of multisource satellite imagery | IEEE Trans. Geosci. Remote Sensing | 309 |
| 7 | Lee et al. | 2010 | Sensing technologies for precision specialty crop production | Comput. Electron. Agric. | 272 |
| 8 | Moulin et al. | 1998 | Combining agricultural crop models and satellite observations: from field to regional scales | Int. J. Remote Sens. | 267 |
| 9 | Carlson et al. | 2012 | Committed carbon emissions, deforestation, and community land conversion from oil palm plantation expansion in West Kalimantan, Indonesia | PNAS | 260 |
| 10 | Ren et al. | 2008 | Regional yield estimation for winter wheat with MODIS-NDVI data in Shandong, China | Int. J. Appl. Earth Obs. Geoinf. | 217 |
| 11 | McNairn and Brisco | 2004 | The application of C-band polarimetric SAR for agriculture: a review | Can. J. Remote Sens. | 208 |
| 12 | Forkuor et al. | 2017 | High Resolution Mapping of Soil Properties Using Remote Sensing Variables in South-Western Burkina Faso: A Comparison of Machine Learning and Multiple Linear Regression Models | PLoS One | 187 |
| 13 | Johnson et al. | 2003 | Mapping vineyard leaf area with multispectral satellite imagery | Comput. Electron. Agric. | 179 |
| 14 | Castillejo-Gonzalez et al. | 2009 | Object- and pixel-based analysis for mapping crops and their agro-environmental associated measures using QuickBird imagery | Comput. Electron. Agric. | 172 |
| 15 | Lopez-Granados | 2011 | Weed detection for site-specific weed management: mapping and real-time approaches | Weed Res. | 168 |
| 16 | Delegido et al. | 2013 | A red-edge spectral index for remote sensing estimation of green LAI over agroecosystems | Eur. J. Agron. | 164 |
| 17 | Muller et al. | 2013 | Comparing the determinants of cropland abandonment in Albania and Romania using boosted regression trees | Agric. Syst. | 163 |
| 18 | Serra et al. | 2003 | Post-classification change detection with data from different sensors: some accuracy considerations | Int. J. Remote Sens. | 162 |
| 19 | Burke and Lobell | 2017 | Satellite-based assessment of yield variation and its determinants in smallholder African systems | PNAS | 156 |
| 20 | Arvor et al. | 2011 | Classification of MODIS EVI time series for crop mapping in the state of Mato Grosso, Brazil | Int. J. Remote Sens. | 155 |

3.1.3 Most used keywords

Following the methodology reported in paragraph 2.3.1, the most frequent stems were retrieved in the publication abstracts driven by a list of relevant keywords categorised into groups (Table A1). All the driving keywords suggested by the authors are retrieved as stems in the WoS dataset, except for "pruning" and "fertilisation" (in this last case the word is stemmed as "fertil" so as not to mislead with the adjective "fertile"). Among the most used keywords (Table A2), the stem "resolut" in the imagery features group is by far the most frequent (518 records), occupying rank #34 in the overall stem count. "Landsat" in combination with the "MODIS" (Moderate Resolution Imaging Spectroradiometer) instrument results in the most widespread satellite platform (290 and 155 records, respectively). The NASA/USGS Landsat program (from Landsat 1 to Landsat 9) provides the longest continuous space-based record of Earth's land in existence while MODIS is a key instrument aboard the Terra and Aqua satellites. Other types of satellite with a stem frequency higher than 20 are "Quickbird", a high-resolution commercial Earth observation satellite, owned by DigitalGlobe, launched in 2001 and reentered in 2015; "Rapideye", a constellation of five satellites owned by Planet, launched in 2008 and deactivated in 2020; and "Terra", the NASA flagship of the Earth Observing System, launched in 1999. The ESA platforms for agricultural applications Sentinel-2 (instrument on board: Multispectral Instrument - MSI) and Sentinel-1 (instrument on board: single C-band synthetic aperture radar - C-SAR) ranked in the second and third position of the satellite category with an overall frequency of 193 ("Sentinel" + "Sentinel-2" + "Sentinel-1"; for frequencies, see Table A2). Being the frequency values quite apart between Landsat and Sentinel stems is certainly due to the Landsat's earlier launch date (1972) in comparison with Sentinel-1 (2014) and Sentinel-2 (2015): this means that Landsat has been broadly used for a much longer time, as it proves its established popularity among researchers but, on the other hand, Sentinel constellation is rapidly spreading for agricultural applications. Regarding crop management, "yield" monitoring/forecast and the "harvest" stage are particularly investigated by researchers while "wheat" is the top-ranking crop. The NDVI (Normalized Difference Vegetation Index) is used as a reference in many studies and results in the most used vegetation index (288 records), in accordance also with the broad use of the "multispectral" sensor. In the image processing topic, the "classification" task achieved the best performance, as well as "radar" among sensors (220 and 134 records, respectively). The keywords belonging to the electromagnetic spectrum group are not present in the top 10 probably also because, like other terms (i.e. those relating to artificial intelligence), researchers followed the general rule of not mentioning acronyms in the abstract.

Analysing the results in sections 3.1.2 and 3.1.3, some considerations on trending topics can be outlined. The classification task on regional and field scales for land use or crop recognition, respectively, seems to be among the most explored research issues. Landsat imagery is widely used. Crop management is focused particularly on yield/monitoring and, among crops, on wheat, which is one of the most important staple crops with 761 million tons produced worldwide [61].

3.1.4 Semantic networks

In the following, authors analyse the semantic network resulting from the validation procedure, described in paragraph 2.2.2, applied to the abstracts of the WoS dataset. To summarise, if a couple of stems appear together in a great number of papers that cannot be explained only by the frequency of the stems in the analysed abstracts and by the verbosity of each abstract, a link connecting those stems is present in the validated projection. The result of the projecting procedure is quite strict [43]: the validated stems are just 751 (over 14294 different stems of the entire dataset) resulting in 5.25% of the original stems.

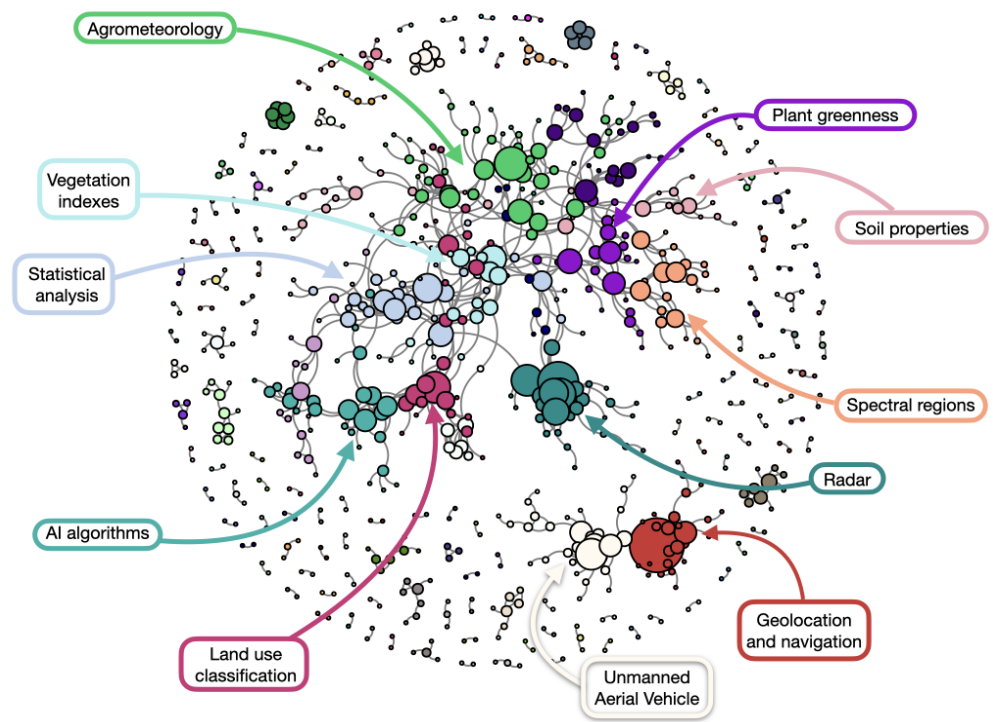

**Figure 8.** WoS dataset semantic network. The arrows mark the communities mentioned in the legend.

As shown by the plot of Figure 8, the network displays a Largest Connected Component (LCC) including nearly 43% of all stems (322 nodes), then a smaller one including 60 nodes, and finally smaller subgraphs made of small cliques. To observe the mesoscale organisation of the system, the Louvain community detection [62] was run. The community detection returns 11 communities (see Table A3 for more details) with more than 10 nodes: 2 of them partition the second largest component. Focus will be placed on the largest connected component and the second largest one since they are the only ones providing a rich relation structure among stems. Nevertheless, regarding the second largest component (bottom right corner of Figure 8), it should be noted that the two communities do not fully fall within the aim of the study. The one on the right represents the "Geolocation and navigation" community (i.e. the first three nodes are "navig", "GNSS" and "posit") encompassing field positioning, robotics and steering. As said before, although these are pivotal applications in agriculture, they do not exploit satellite imagery and then are excluded from the study's aim. The other community is labelled as "Unmanned Aerial Vehicle" (UAV) and contains stems related to drones, dedicated sensors ("RGB" and "lidar") and typical UAV-derived imagery products as "DEM" (Digital Elevation Model). Many of the publications retrieved in the dataset dealing with UAVs seem to somehow mention satellites merely as another RS platform, often to enhance the advantages of drones. Data comparison between satellite and UAV platforms for calibration/validation or upscaling/downscaling processes seems to be quite rare considering both the absence of the keyword "interoperability" in the validated network and the fact that this community is part of a connected component different from LCC. However, the two communities are closely related and this agrees with their integrated involvement in PA activities on field/parcel scales.

Shifting to LCC, it can be noted that the largest community (47 nods) labelled as "Agrometeorology", occupies a central position within the network and it is well connected with numerous links to almost all the others. This community collects terms related to agrometeorological parameters and retrieving methods influencing crop growth; therefore, it is at the centre of the network as the weather conditions are the main driving factors of agricultural activity. Two other large communities with the same number of nodes (31) are "Radar" and "Soil properties". These occupy a peripheral space of LCC and do not appear to be closely connected with other communities. The "Radar" community presents very numerous nodes that highlight the main radar platforms used in the agricultural sector (Radarsat-2, Sentinel-1 and ALOS with the Palsar instrument). Instead, the "Soil properties" community has a tree structure, starting from stem nodes relating to the soil organic content and electrical conductivity. The latter in turn branch into indications about texture ("clay"), salinity ("salin") and nutrient availability ("pH", "phosphorus").

In the bottom part of Figure 7, two communities that are strictly linked can be noticed: "Land use classification" and "Artificial Intelligence (AI) algorithms". The latter comprehends stems from both machine learning ("svm", "rf") and deep learning ("ann", "cnn") algorithms, which are often used for land use classification tasks. It is worth to be mentioned that in the "Land use classification" community occurs a missed identification of stems, as presented in paragraph 2.2.1: "classif" and "classifi" (Table A3) refer to the same concepts, as they are stems of "classification", "classify" and "classifier", and their conjugations. The retrieving of chlorophyll and nitrogen content is the main core of the "Plant greenness" community, which is connected with the "Spectral regions", where portions of the electromagnetic spectrum typically employed in agricultural tasks are reported ("red", "infrar", "nir"). Finally, despite its modest size (22 nodes), at the centre of LCC, there is the "Vegetational indexes" community: the large number of links with other communities is justified by the broad use of vegetation indexes in processing satellite imagery for agricultural research activities.

Focusing on Sentinel-2 satellite, it must be highlighted that the community which includes this platform is not reported (< 10 nodes). However, due to the role that Sentinel-2 plays in agricultural applications, the authors analysed its community and compared it with the one encompassing Sentinel-1 ("Radar" community). The stems of the two communities show that in the case of Sentinel-1, the scientific discourse revolves around the acquisition of radar imagery ("apertur", "backscatt", "vv") while for Sentinel-2 the researchers' focus is pointed on the extraction of informative data from images, often through machine learning and deep learning techniques ("svm", "rf", "neural", "convolut").

To consider even arguments connecting different fields, as represented in terms of communities of stems, the node betweenness is considered, i.e. an index counting the number of shortest paths that pass through each node [35]. Table 2 reports the betweenness for both overall network stems and stems related to the driving keywords selected by the authors. The values are normalised such that the node with the greatest betweenness has a value of 1 and the one with the lowest a value of 0. As it could be expected, the terms of the overall network connecting the various communities are mainly generic and related to image analysis ("coeffici", "index", "correl", "classif") or other sorts of ubiquitous in the use of satellite imagery in agriculture ("moistur", "soil", "leaf").

All selected stems show lower values of betweenness and it is not surprising: all of them are more specific, therefore they are quite central in their community, but are limitedly connected outside.

**Table 2.** Betweenness of stems in WoS semantic network. The top 10 stems, in terms of betweenness, within the entire semantic network (left columns) and the list of stems selected by the authors (right columns) are reported.

| Overall top 10 | | Selected top 10 | |
|---|---|---|---|
| *Stem* | *Betweenness* | *Stem* | *Betweenness* |
| coeffici | 1 | modi | 0.346 |
| moistur | 0.838 | thermal | 0.332 |
| soil | 0.779 | yield | 0.269 |
| leaf | 0.73 | radar | 0.26 |
| index | 0.662 | wheat | 0.174 |
| kappa | 0.646 | evi | 0.136 |
| correl | 0.565 | resolut | 0.125 |
| backscatt | 0.547 | hyperspectr | 0.1 |
| temperatur | 0.535 | ndvi | 0.075 |
| classif | 0.512 | landsat | 0.061 |

*3.2 Twitter dataset*

3.2.1 Descriptive statistics

Similarly to what was observed in Figure 2, even on Twitter the volume of messages targeting the use of satellite imagery in agriculture experienced a rapid increase since the foundation of the platform, as can be observed from Figure 9. Nevertheless, differently from the WoS dataset, there is an evident decrease in 2021 that is not present in the WoS dataset. Probably, this decline occurs since Twitter is intensively used to advertise activities like workshops, conferences and live events: due to the COVID-19 pandemic, in 2021 most of those activities were cancelled, if they did not undergo strong limitations.

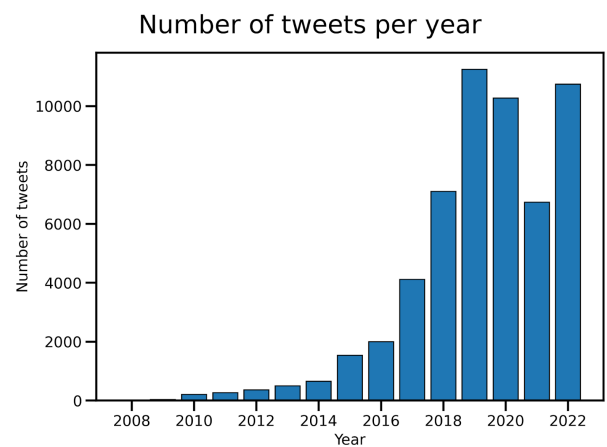

**Figure 8.** Number of tweets per year (2008 - mid2022).

Analogously to Figure 7, in Figure 10 some distributions regarding the Twitter dataset are represented. In particular the number of retweets (i.e. sharings) per message (Figure 10a) and the number of original messages (i.e. tweets) per account (Figure 10b). In Figure 10, panels (a) and (b) are analogous to the number of citations per paper (Figure 7a) and the number of papers per author (left panel, Figure 7b). As in the case of the WoS dataset, even the distributions displayed in Figure 10 are power laws, thus providing additional confirmations about the "adjacent possible" interpretation formulated in [46].

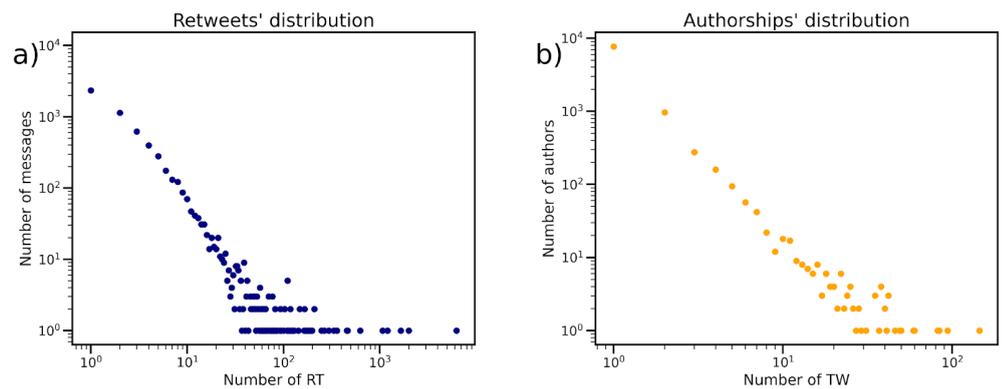

**Figure 10**. Retweets and authorships' distributions throughout the Twitter dataset (2008 - mid2022).

3.2.2 Semantic Network

Similarly to the analysis of section 3.1.4, the authors analysed the semantic network resulting from the validation procedure, applied to the Twitter posts. Firstly, tweets containing original contents, i.e. all messages that are not retweets, were selected. Then, each message was reduced in stems (including hashtags and emoticons. Emoticons were not stemmed, obviously, but were considered in the bipartite network). Finally, the procedure involved building a bipartite network of tweet ids (i.e. the unique identifier of each tweet message) and stems (as obtained with the procedure described in paragraph 2.2.1) and validating the projection on the layer of stems, using BiCM as a benchmark.

The resulting network includes 4767 nodes and 17843 links. In this case, the network is composed of a Largest Connected Component (pictorially represented in Figure 11), including nearly 72.5% of the nodes of the entire network.

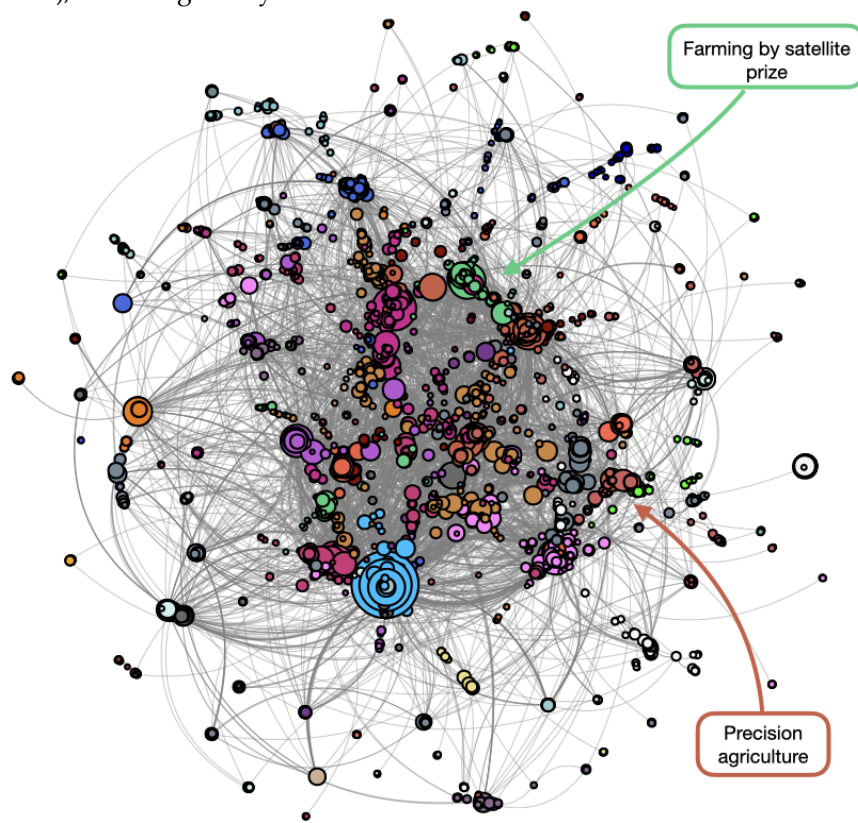

**Figure 11.** Twitter dataset semantic network. The arrows mark the communities mentioned in the legend. Other communities are related to arguments different from the result of academic research

in the use of satellite imagery in agriculture and therefore are neglected. As it can be seen, academic achievements represent a small portion of the semantic network that, instead, is mostly populated by arguments related to the presentation of new products by private companies.

To detect the mesoscale organisation of the network, the Louvain algorithm for community detection [62] was run: the algorithm returned 329 communities, but only 14 have more than 100 nodes. Indeed, the communities found are much larger than the one observed in the WoS dataset, i.e. the most numerous one accounts for 255 nodes. As the communities become larger, their identification in terms of main topics suitable for research purposes becomes tougher, since many subtopics contribute to the same community.

The volume of messages in the network is given primarily by the massive number of tweets posted by private companies that use Twitter as a showcase to promote their activities. The enterprise's core business includes the satellite and space industry in general but also specific activities such as providing satellite data through dedicated web platforms directly to the farmers, even for developing nations. Companies that specifically deal with the agricultural sector are committed to making new technologies (not just satellite data) more accessible to farmers. A secondary high volume of messages come from national space agencies that announce the successful launch of new satellites (i.e. the hyperspectral satellite Prisma through the Vega launcher in 2019).

Moving to agriculture-oriented topics, only two communities worthy of being commented on for the aim of the study were found in the network, labelled as "Farming by satellite prize" (220 nodes) and "Precision agriculture" (129 nodes). The "Farming by satellite prize" community bears witness to the intense technology transfer and communication activity put in place by the European institutions about their space programs. It comprehends among the most significant stems "copernicus", "egno" and "fbs prize". The prize aims to support young innovators in delivering applications and services based on Galileo, EGNOS (European Geostationary Navigation Overlay System) and Copernicus that will contribute to the evolution of agriculture. The "Precision farming" community emerges as the only cluster with technical stems but, regarding the use of satellite images in agriculture, they can be considered quite generic. The stems "water", "veget", "organicfarm" and "crophealth" represent the community core with a frequency of 34, 31, 29 and 26, respectively.

As in paragraph 3.1.4, to consider even arguments connecting different fields, the node betweenness for both overall network stems and stems selected through the driving keywords was considered (Table 3). The stems in the overall network with the highest betweenness values ("iot", "co2", "ecosensenow") refer to generic terms outside the use of satellite imagery in agriculture. This is probably due to the diversified activities carried on by private companies that animate the debate on Twitter in this sector. Switching on the selected stems, the betweenness has low values except for "radar" and "ai". In the first case, the good value of betweenness is related to the extensive advertising that ISRO made about its radar satellite launches. Since the stem "ai" is often present in communities driven by private companies, the betweenness of AI techniques could derive from business solutions outside the topic under investigation.

**Table 3.** Betweenness of stems in the Twitter semantic network. The top 10 stems with the overall greatest betweenness (left columns) and the top 10 selected stems (right columns) are reported.

| Overall top 10 | | Selected top 10 | |
| --- | --- | --- | --- |
| *Stem* | *Betweenness* | *Stem* | *Betweenness* |
| iot | 1.000 | radar | 0.400 |
| co2 | 0.933 | ai | 0.303 |
| ecosensenow | 0.903 | yield | 0.164 |
| imag | 0.682 | thermal | 0.101 |
| earth | 0.669 | resolut | 0.092 |
| fire | 0.641 | landsat | 0.055 |
| emiss | 0.633 | rice | 0.044 |
| launch | 0.628 | pest | 0.043 |
| farm | 0.627 | harvest | 0.040 |
| prize | 0.621 | aqua | 0.039 |

**4. Conclusions**

To the best of the authors' knowledge, this is the first study performing an informetric analysis on the use of satellite imagery in agriculture leveraging both a scientific literature dataset and a non-academic information source.

In the Introduction, 5 research questions (RQ) were proposed, to better organise the presentation of the results. The first 3 ones regarded the analysis of the scientific literature related to remote sensing in precision agriculture, while the following 2 tackled the presence of the same subject on Twitter. To highlight non-trivial relations and the evolution of satellite imagery in agriculture within scientific research (i.e. answering the first 3 RQs), papers were downloaded from the Web of Science by selecting "agriculture" (and its conjugations) and "satellites" in the abstract, title and keywords. After a deep data cleaning, necessary to avoid papers related to distant fields, descriptive statistics and the semantic network of stems were extracted.

First, the performances of countries, journals and the most influential papers were analysed, answering RQ1. Regarding the evolution of the scientific field, a rapid growth in the production of papers was observed during the years, due, essentially, to the evolution of technology for satellite imagery access and their automated analysis. Furthermore, it is noteworthy that the predominance of the USA in this field has been challenged by China in the very last years, both in terms of publications and in employed researchers. Nevertheless, the research impact of US publications remains much higher than those of China.

When analysing the most frequent scientific journals, authors observed that the ones related to long-established editors (Elsevier, IEEE) are challenged by more recent ones (MDPI, PLOS). Nevertheless, the study of the distribution of citations showed that long-established editors can attract the greatest attention.

Mainstream topics were retrieved from the analysis of the most cited papers and most frequent abstract keywords selected through a tailored list. Hence, it follows that radar and multispectral sensors are broadly used, and among the latter in particular MODIS, a key instrument aboard the Terra and Aqua NASA satellites. Another milestone of the agricultural research task is represented by the computation of the NDVI vegetation index through satellite imagery. Among crops, wheat often plays a key role in different studies with a focus on yield forecasting.

To answer RQs 2 and 3, i.e. the analysis of the terminology used in the scientific literature and the relation among the various terms (RQ2) and the inference of emerging trends (RQ3), the semantic network, as extracted from the abstract of the various papers, was then analysed. More in detail, all words from the abstract of each paper were collected, stop words and punctuations were removed. Finally, all retrieved words were "stemmed": the rationale was to focus on the concepts and not on the single word; therefore, different conjugations of the same word were reduced to the same stem.

Finally, the non-trivial co-occurrences of stems in papers' abstracts were filtered via an entropy-based approach. In a nutshell, this approach resides in comparing the observations to a benchmark that is maximally random but for some information of the real system.

The result is a network of stems that are connected if they are significantly used together in the various abstracts. Thus, it is possible to detect different communities, related to the various subfields of the research.

From the network analysis, it emerges the large size of the "Agrometeorology" community, confirming the importance of RS retrieval of meteorological parameters in the agricultural sector, also as a base for further studies. Although with smaller dimensions, it is worth to be noted that the "Vegetation indexes" community is well connected with others. Moreover, some indexes (i.e. EVI and NDVI) included in this community have low values of betweenness; nodes with low values of betweenness are well-connected in their community, but relatively loosely linked to nodes in other communities.

Integrating all information in a coherent view, it is remarkable how the most cited papers are quite recent: for instance, the most cited paper is only 11 years old. In this sense, it seems that the "first mover advantage", i.e. the fact that older papers gather more citations just for the fact of being in the literature for a longer time, is not relevant. Instead, probably due to the rapid growth of the present field of research, older papers may have become outdated quite soon, and therefore have not exploited their time advantage. On the other hand, the network of validated stems highlights the presence of emerging topics. Analysing the stems retrieved in recently published papers, classification task for land use assessment and crop recognition stands out as a pervasive and cutting-edge topic, together with the use of hyperspectral sensors and RS retrieving of plant responses to environmental cues (i.e. fluorescence).

Nonetheless, the academic contribution is just one facet of the impact of RS in PA. To exploit even the impact of these novel tools on practitioners, data from online social networks – in particular Twitter for its frequency among private companies and for its data availability – were also analysed: in particular, the research aimed to infer the different purposes in the online debate on remote sensing in precision agriculture (RQ4) and the analysis of the dynamic and the development of the various topics (RQ5). In downloading the data from Twitter API, the search queries mirrored the one used for the WoS data retrieval to provide an easy comparison between results. Indeed, some descriptive statistics are quite similar (such as the retweet/citations distributions), but there are also evident differences, such as a clear decline in the number of tweets in 2021: such behaviour, that is not present in the academic production, is probably due to the fact that companies on Twitter often advertise public events, that were cancelled or postponed during the COVID-19. One of the main findings of the study is that Twitter is largely employed by private companies and national space agencies for informational and promotional purposes. On the contrary, it does not seem to be very popular among researchers to share and spread scientific findings outside academia.

From a scientific point of view, future challenges concern the availability of satellite imagery with high resolution and shortened revisit times. The processing bottleneck related to the increasing amount of data may increase but all the agricultural RS tasks will benefit from these new achievements which will pave the way for more accurate or new applications. In this sense, researchers could focus on developing new models (i.e. crop growth) embedding various types of satellite data or on associating plant spectral signatures to crop features or status, such as variety or disease-induced stress.

From the bibliometric point of view, the present analysis shows a novel and promising application of entropy-based null-models for the detection of semantic networks. Similar procedures have been applied in Twitter datasets [33,63], where the workflow was limited to the usage of hashtags, while in this study, the attention is on the entire abstract, in the case of the WoS dataset, or on the entire text of the message, in the case of the Twitter dataset.

Regarding the bibliometric analysis of remote sensing in precision agriculture, while this research focuses on Twitter and WoS, other data sources could complement the analysis. For instance, regarding the academic facet, Google Scholar includes other documents from grey literature, e.g. Ph.D. and Master's thesis and may reinforce or open up new clusters of stems in the semantic network. However, due to the tendency of Google Scholar to include different data, for instance not necessarily peer-reviewed, the risk of adding uninformative noise to the dataset is quite high. Even focusing on a "cleaned" source as Web of Science, the presence of noise was quite strong and adding grey literature may introduce an extra source of noise. Nevertheless, analysing data from Google Scholar too could be of interest, to highlight similarities and differences with WoS. Analogously, regarding the online social network side, it could be interesting to study even the data from other social networks, such as Facebook. While, in this case, the data availability is quite limited, as well as the detail of the data about users, it could represent an interesting source to be investigated.

The main aim of PA research is to provide more efficient tools to lower the impact on natural resources and, at the same time, increase profits for farmers. In this study, a limited presence of academia promoting its scientific advancements in online social networks is highlighted. Online social networks represent a novel and unprecedented way to target practitioners and privates, therefore if academia will not exploit this opportunity, achievements in the use of satellite imagery will struggle to reach a wider audience and their effectiveness may be subject to limitations.

**Supplementary Materials:** The following supporting information can be downloaded at: www.mdpi.com/xxx/s1.

**Author Contributions:** Conceptualization, R.D. and F.S.; methodology, R.D. and F.S.; software, F.S.; validation, R.D.; formal analysis, F.S.; investigation, R.D.; data curation, F.S.; writing—original draft preparation, R.D. and F.S.; writing—review and editing, R.D. and F.S.; visualization, R.D. and F.S.; supervision, R.D. All authors have read and agreed to the published version of the manuscript.

**Funding:** This research received no external funding.

**Data Availability Statement:** Twitter data can be downloaded from the following link (https://drive.google.com/file/d/1sM4HYLwZD_WC5jYn80M6t-qugsPXU5iv/view?usp=sharing) as the list of the tweet ids used in the analysis, in agreement with the Twitter policy. (https://developer.twitter.com/en/developer-terms/more-on-restricted-use-cases). Web of Science data are proprietary and cannot be shared without the permission of WoS.

**Conflicts of Interest:** The authors declare no conflict of interest.

**Appendix A**

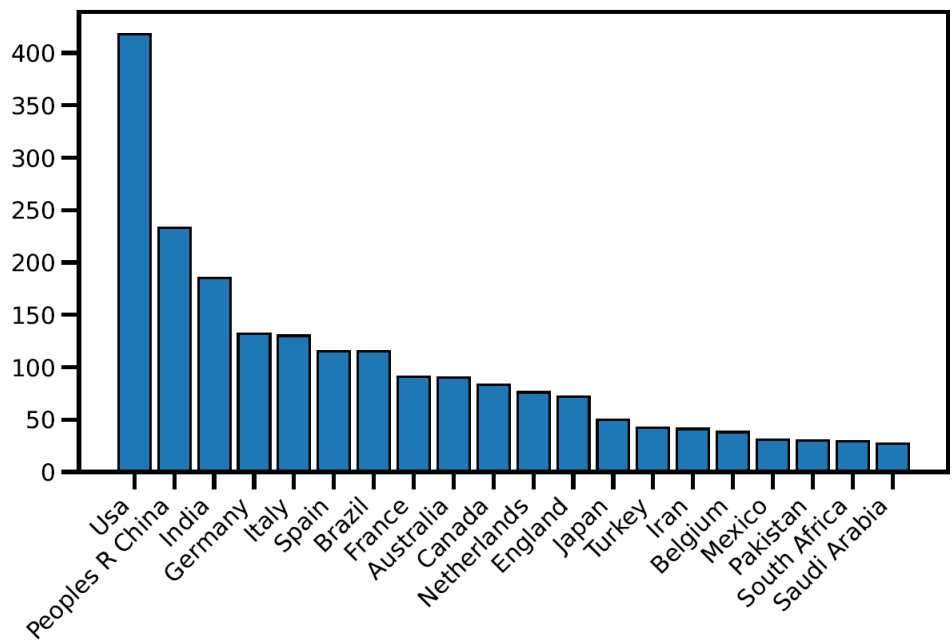

**Figure A1.** Top 20 countries in terms of publication (1977 - mid2022).

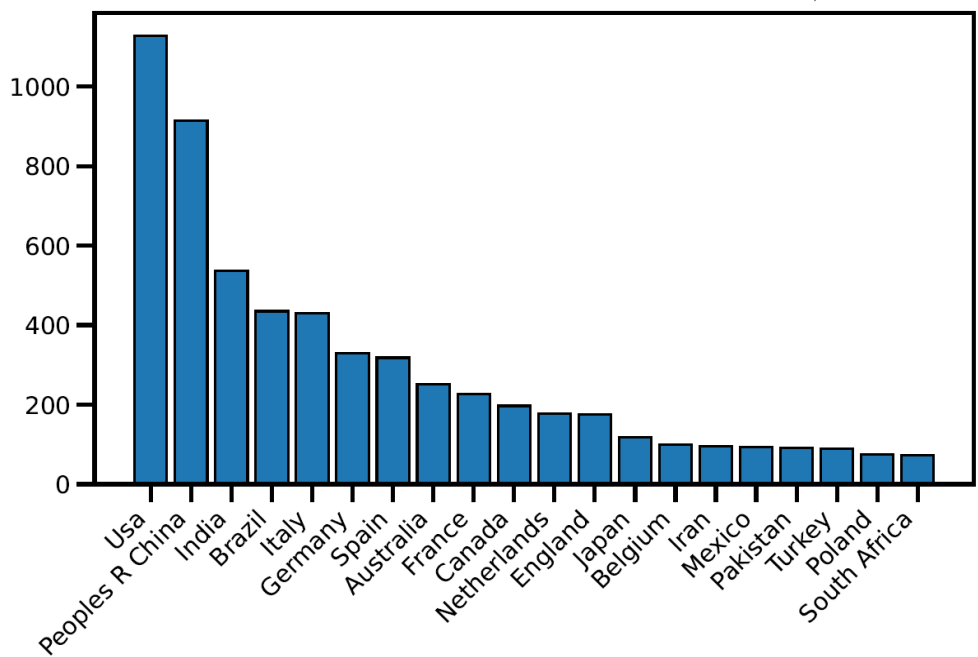

**Figure A2**. Top 20 countries in terms of researchers (1977 - mid2022).

**Table A1.** Driving list of relevant keywords categorised into groups and used for agricultural keywords selection.

| Satellite platform/ instrument | Sensor type | Electromagnetic spectrum | Vegetation index | Imagery features | Image processing | Crop type | Crop management |
|---|---|---|---|---|---|---|---|
| ALOS | Hyperspectral | NIR | EVI | Multitemporal | ⁺AI | Cereals | Disease |
| Aqua | Multispectral | SWIR | GCI | Resolution | Classification | Barley | Fertilisation |
| Gaofen | Radar | TIR | GNDVI | Revisit time | Fusion | Corn | Harvest |
| Landsat | RGB | VIS | LAI | | ⁺DL | Maize | Pest |
| PlanetScope | Thermal | | MSAVI | | Interoperability | Rice | Phenology |
| Pleiades | | | NDRE | | ⁺ML | Wheat | Pruning |
| Prisma | | | NDVI | | Recognition | Banana | Tillage |
| QuickBird | | | NDWI | | | Oil palm | Sowing |
| RadarSAT | | | RECI | | | Soybean | Weed |
| RapidEye | | | SAVI | | | Sugarcane | Yield |
| Sentinel | | | VARI | | | Potato | |
| Terra | | | | | | Tomato | |
| MODIS* | | | | | | Orchards | |
| | | | | | | Olive | |
| | | | | | | Vineyard | |

* instrument
⁺ AI = Artificial intelligence; DL = Deep learning; ML = Machine learning

Table A2. Stem, frequency and ranking of agricultural keywords selected through the driving list.

| Research field | Keyword | Stem | Frequency | Ranking |
|---|---|---|---|---|
| imagery_features | resolution | resolut | 518 | 34 |
| crop_management | yield | yield | 380 | 65 |
| satellites | landsat | landsat | 290 | 107 |
| vegetation_index | ndvi | ndvi | 288 | 108 |
| image_processing | classification | classifi | 220 | 156 |
| crop_type | wheat | wheat | 178 | 217 |
| satellites | modis | modi | 155 | 268 |
| sensors | radar | radar | 134 | 312 |
| sensors | multispectral | multispectr | 128 | 329 |
| crop_management | harvest | harvest | 119 | 353 |
| vegetation_index | vari | vari | 115 | 363 |
| crop_type | rice | rice | 109 | 380 |
| crop_management | phenology | phenolog | 103 | 416 |
| crop_type | maize | maiz | 94 | 451 |
| vegetation_index | lai | lai | 82 | 506 |
| sensors | hyperspectral | hyperspectr | 79 | 529 |
| sensors | thermal | thermal | 72 | 576 |
| crop_type | corn | corn | 70 | 591 |
| vegetation_index | evi | evi | 58 | 672 |
| crop_type | soybean | soybean | 58 | 674 |
| image_processing | fusion | fusion | 49 | 764 |
| crop_management | disease | diseas | 46 | 794 |
| imagery_features | multitemporal | multitempor | 42 | 866 |
| crop_management | weed | weed | 42 | 867 |
| satellites | sentinel | sentinel | 39 | 891 |
| em_spectrum | vis | vis | 39 | 905 |
| crop_type | cereals | cereal | 36 | 950 |
| crop_type | potato | potato | 33 | 989 |
| crop_management | pest | pest | 32 | 1017 |
| em_spectrum | nir | nir | 30 | 1064 |
| crop_type | barley | barley | 30 | 1057 |
| crop_type | sugarcane | sugarcan | 30 | 1076 |
| crop_management | sowing | sow | 30 | 1060 |
| crop_management | tillage | tillag | 27 | 1135 |
| satellites | quickbird | quickbird | 26 | 1164 |
| satellites | rapideye | rapidey | 26 | 1169 |
| imagery_features | revisit time | revisit | 23 | 1271 |
| crop_type | orchards | orchard | 23 | 1242 |
| vegetation_index | ndwi | ndwi | 22 | 1307 |
| satellites | terra | terra | 22 | 1300 |
| vegetation_index | savi | savi | 21 | 1337 |
| image_processing | ml | ml | 19 | 1444 |
| image_processing | recognition | recognit | 19 | 1428 |
| sensors | rgb | rgb | 18 | 1470 |
| crop_type | olive | oliv | 17 | 1526 |
| crop_type | vineyard | vineyard | 17 | 1517 |
| crop_type | oil palm | palm | 16 | 1594 |
| satellites | alos | alo | 15 | 1612 |
| em_spectrum | tir | tir | 14 | 1686 |
| vegetation_index | gndvi | gndvi | 13 | 1780 |
| satellites | aqua | aqua | 12 | 1841 |
| vegetation_index | ndre | ndre | 11 | 1986 |
| satellites | radarsat | radarsat | 9 | 2164 |
| em_spectrum | swir | swir | 9 | 2171 |
| satellites | planetscope | planetscop | 8 | 2350 |
| crop_type | tomato | tomato | 8 | 2319 |
| vegetation_index | msavi | msavi | 5 | 2944 |
| image_processing | interoperability | interoper | 5 | 2869 |
| image_processing | dl | dl | 4 | 3368 |
| crop_type | banana | banana | 4 | 3451 |
| image_processing | ai | ai | 3 | 4021 |
| satellites | gaofen | gaofen | 2 | 4403 |
| satellites | prisma | prism | 2 | 5033 |
| vegetation_index | gci | gci | 1 | 13093 |
| vegetation_index | reci | reci | 1 | 8925 |
| satellites | pleiades | pleiad | 1 | 6140 |

**Table A3.** Overview of the community features: label, number of nodes and stems with frequency.

| Community label | Agro-meteorology | Statistical analysis | Radar | Soil properties | Land use classification | Geolocation and navigation | Unmanned Aerial Vehicle | AI algorithms | Spectral regions | Vegetation indexes | Plant greenness |
|---|---|---|---|---|---|---|---|---|---|---|---|
| Number of nodes | 47 | 37 | 33 | 31 | 31 | 30 | 30 | 27 | 24 | 22 | 22 |
| Community color | | | | | | | | | | | |
| Stems and frequency | evapotranspir 14 | rmse 12 | radar 17 | carbon 6 | classif 13 | navig 23 | unman 13 | machin 9 | band 8 | index 11 | leaf 10 |
| | temperature 10 | coeffici 10 | apertur 15 | organ 6 | classifi 8 | gnss 11 | aerial 9 | vector 8 | red 8 | modi 8 | nitrogen 9 |
| | water 8 | regress 9 | sar 13 | soil 6 | hulc 8 | posit 9 | uav 9 | learn 8 | reflect 8 | veget 7 | chlorophyl 7 |
| | precipit8 8 | squar 9 | backscatt 13 | electr 5 | likelihood 7 | kinemat 8 | vehicl 9 | random 7 | infrar 7 | ndvi 7 | n 6 |
| | thermal 7 | moistur 8 | synthet 12 | clay 4 | kappa 7 | receiv 5 | precis 6 | rf 6 | hyperspectr 7 | evi 6 | content 5 |
| | balance 7 | correl 7 | polar 11 | induct 3 | accuraci 6 | rtk 5 | flight 4 | svm 6 | nn 4 | spectroradiomet 5 | -1 4 |
| | surface 7 | error 7 | vv 11 | salin 3 | urban 6 | real-tim 4 | platform 4 | neural 6 | nir 4 | resolut 5 | solar-induc 4 |
| | drought 7 | predict 6 | hh 7 | ec 3 | chang 5 | system 4 | camera 4 | network 5 | near-infrar 4 | normal 4 | kg 4 |
| | et 6 | root 5 | c-band 7 | eca 3 | class 5 | global 4 | dem 3 | convolut 5 | spectral 3 | moder 4 | canopi 3 |
| | energi 6 | mean 5 | l-band 7 | matter 3 | cover 5 | signal 3 | digit 3 | deep 5 | spectra 3 | differ 3 | fertil 3 |
| | rainfall 5 | least 5 | radarsat-2 6 | soc 2 | overal 4 | speed 3 | elev 3 | forest 5 | green 2 | spatial 3 | fluoresc 3 |
| | tmm 5 | plsr 4 | microwav 6 | phosphorus 2 | maximum 4 | geograph 3 | uav-bas 3 | decis 4 | visibl 2 | indic 3 | lai 3 |
| | meteorology 4 | absolut 4 | optic 5 | ph 2 | built-up 3 | gis 2 | height 3 | support 4 | tir 2 | terra 2 | sif 2 |
| | 1st 4 | r-2 4 | palsar 5 | appar 2 | supervis 3 | path 2 | rgb 3 | artifici 4 | blue 2 | imag 2 | rate 2 |
| | emiss 4 | retriev 3 | sentinel-1 5 | eddi 2 | bodi 3 | dgps 2 | topographi 2 | tropic 3 | red-edg 2 | high 2 | nutrient 2 |
| | actual 3 | passiv 3 | polarimetr 5 | ecosystem 2 | land 3 | real 2 | strm 2 | ann 3 | hyperion 2 | savi 2 | ha 2 |
| | sebal 3 | ordinary 3 | vh 3 | depth 2 | sprawl 2 | autonom 2 | man 2 | train 2 | shortwav 2 | low 1 | biomass 2 |
| | vci 3 | partial 3 | alo 3 | fao 1 | nlc 2 | steer 1 | altitud 2 | algorithm 2 | swir 1 | ndvi 1 | biophys 2 |
| | climat 3 | valid 3 | angl 3 | cm 1 | use/land 2 | rs 1 | lidar 2 | deforest 2 | airborn 1 | distribut 1 | photosynthesi 1 |
| | gaug 3 | model 2 | s2 2 | servic 1 | increas 2 | robot 1 | farmer 2 | cnn 2 | multispectr 1 | aqua 1 | absorpt 1 |

The first 20 stems per community are reported.